\documentclass[acmsmall,screen,nonacm]{acmart}

\usepackage{soul}
\usepackage{xspace}
\usepackage[]{hyperref}
\usepackage{comment}
\usepackage{cprotect}
\usepackage{multirow}
\usepackage{tabularx}
\usepackage{xcolor}
\usepackage[inline]{enumitem}
\usepackage{listings}
\usepackage{subcaption}
\usepackage{makecell}
\usepackage{makecell}
\usepackage{fancyvrb,newverbs}
\usepackage{booktabs}
\usepackage{anyfontsize}
\usepackage{hhline}
\usepackage{boldline}
\usepackage{array}
\usepackage{arydshln}
\usepackage{float}
\usepackage{cprotect}

\newcommand{\google}{Google\xspace}

\newcommand{\numberOfChanges}{38,156\xspace}
\newcommand{\totalAutomatedCommits}{31,984\xspace}
\newcommand{\automatedCommitFraction}{83.82\%\xspace}
\newcommand{\automatedCommitLoCFraction}{14.15\%\xspace}

\newcommand{\categoryCommitFractionConfig}{84\%\xspace}
\newcommand{\categoryLoCFractionConfig}{19\%\xspace}
\newcommand{\categoryLoCFractionTooling}{17\%\xspace}

\lstdefinelanguage{code}{
  basicstyle=\fontsize{8}{8}\selectfont\ttfamily,
  backgroundcolor=\color{lightgray!8},
  morecomment=[f][\diffbg{red!20}]-,
  morecomment=[f][\diffbg{green!20}]+,
  morecomment=[f][\textit]{@@},
}

\begin{document}

\title{Instruction Set Migration at Warehouse Scale}

\author{Eric Christopher, Kevin Crossan, Wolff Dobson, Chris Kennelly, Drew Lewis, Kun Lin, Martin Maas, Parthasarathy Ranganathan, Emma Rapati, Brian Yang}

\renewcommand{\shortauthors}{}

\affiliation{
  \institution{Google}
  \country{USA}
}

\authorsaddresses{}

\begin{abstract}
Migrating codebases from one instruction set architecture (ISA) to another is a major engineering challenge. A recent example is the adoption of Arm (in addition to x86) across the major Cloud hyperscalers. Yet, this problem has seen limited attention by the academic community. Most work has focused on static and dynamic binary translation, and the traditional conventional wisdom has been that this is the primary challenge.

In this paper, we show that this is no longer the case. Modern ISA migrations can often build on a robust open-source ecosystem, making it possible to recompile all relevant software from scratch. This introduces a new and multifaceted set of challenges, which are different from binary translation.

By analyzing a large-scale migration from x86 to Arm at \google, spanning almost 40,000 code commits, we derive a taxonomy of tasks involved in ISA migration. We show how \google automated many of the steps involved, and demonstrate how AI can play a major role in automatically addressing these tasks. We identify tasks that remain challenging and highlight research challenges that warrant further attention.
\end{abstract}

\maketitle

\section{Introduction}
\label{sec:introduction}

Migrating large codebases to a new Instruction Set Architecture (ISA) is a major engineering challenge. Examples include Apple’s migration from PowerPC to x86 and later to Arm \cite{apple_transition}, as well as the adoption of Arm by major hyperscalers (such as Amazon, Google, Microsoft). While there are anecdotal claims regarding the complexity and efforts required for such migrations \cite{825694,MulticoreWare2024Porting,10.1145/945131.945155}, to our knowledge, there is no systematic analysis of what these ISA migrations entail, and how they are impacted by modern technologies such as improved software engineering tools and artificial intelligence (AI). In this paper, we perform such a systematic analysis for the migration of a multi-billion line codebase from x86 to Arm at \google.

Historically, the conventional wisdom has been that the biggest challenge in ISA migration involves translating machine code between ISAs \cite{825694,10.1145/945131.945155}. Correspondingly, there has been a significant amount of work on static \cite{10.1145/2380403.2380419} and dynamic \cite{10.1145/3567955.3567962} binary translation that automatically rewrites binaries compiled for one ISA to another. Binary translation was the main problem when software was distributed as binaries and source code was not usually available. However, modern ISAs are generally well-supported in upstream compilers, runtime libraries, and the Linux kernel. As a result, modern compilers mostly ``just work'' for a new ISA, and previous ISA migrations have smoothed the path to packages supporting cross-compilation by default. For example, 98\% of Debian packages build for RISC-V, although it only became an official Debian architecture in 2023 \cite{debianRISCVDebian}.

Perhaps surprisingly, this does not mean that ISA migration is no longer a challenge. While code translation is not the main issue anymore, we find that modern ISA migration involves many usually-simple, repetitive, automatable tasks such as updating build scripts or fixing floating-point issues, which AI can increasingly facilitate. In this paper, we analyze a large-scale ISA migration at \google that added Arm support alongside x86. We focus on the following research questions:

\begin{enumerate}
\item What are the tasks that are involved in a modern ISA migration?
\item Which tasks can be automated and how can modern AI help?
\item Which tasks are difficult and are good targets for future research?
\end{enumerate}

\noindent To answer these questions, we provide what we believe is a first-ever detailed breakdown and taxonomy of large-scale ISA migration tasks. Using state-of-the-art LLMs, we analyze and categorize a corpus of \numberOfChanges commits that constitute our real-world migration. We quantitatively evaluate the capability of current tools, including AI models, to perform these tasks automatically. We systematically identify the strengths and weaknesses of current automated tools, and highlight areas of future work and improvement. We believe that this work highlights research opportunities for the academic community and revisits long-standing assumptions around ISA migrations.

Specifically, we contribute the following insights:
\begin{enumerate*}[label=\emph{\arabic*}), itemjoin={{; }}]
\item The complexity of ISA migrations is not in code translation but involves a number of different tasks, many related to rewriting BUILD and configuration files
\item Many of these tasks are highly automatable
\item Many of the tasks that are not automatable only need to be performed once when going from a single ISA to multiarch
\item Of the remaining tasks, many can be performed by modern AI, but some challenges remain.
\end{enumerate*}

\section{Background \& Related Work}

There are a number of reasons why organizations have performed large-scale ISA migrations. First, many ISAs have gone extinct over the years (e.g., Alpha, MIPS, SPARC, Itanium, VAX). Second, with the adoption of Android and iOS, more codebases were ported to Arm to be used in mobile applications. Third, Apple Macs went through successful migrations from PowerPC to x86, and most recently from x86 to Arm to support custom Apple Silicon. Finally, major cloud hyperscalers have been migrating large codebases from x86 to Arm as well.

In this context, migration does not only refer to the act of getting software to build on a new architecture but to reaching parity in terms of performance, security and stability. The most closely-related work in the academic community falls into two categories.

First, there is a significant amount of work on static and dynamic binary translation from one ISA to another \cite{10.1145/3567955.3567962,10.1145/2380403.2380419}. Static or dynamic binary translators can serve as a bridge to handle the long tail of a software ecosystem, but eventually they must be phased out to avoid carrying forward technical debt (e.g., Apple will actively phase out its Rosetta dynamic binary translation system in 2027 \cite{appleRosetta}). In cloud computing, where compute is commoditized, developers seek as much efficiency as possible, and recompiling onto the new ISA is the best route to maximize the compiler's options for performance. We thus forewent binary translation altogether in our deployment.

Second, there is a significant amount of work on automatically applying edits to code, such as for performance optimization \cite{lin2025ecollmdrivenefficientcode}, fixing security issues \cite{52980}, or correcting bugs \cite{10.1145/3360585}. As we will see, these are common tasks that are part of a successful ISA migration.

\section{\google's x86 to Arm Journey}

We now analyze \google's multi-year effort to port a substantial portion of \google's server application ecosystem from x86 to Arm, enabling simultaneous support for both. We start by describing Google’s environment and provide a step-by-step analysis of our ISA migration.

\subsection{\google's Software Ecosystem}

\google's codebase is organized as a monorepo containing billions of lines of code \cite{45424}. Individual applications and libraries reside in various directories. These folders also contain metadata files, e.g., to indicate code owners or configure continuous integration (CI) testing \cite{winters2020software}.

\paragraph{Building with Bazel}

Builds use Bazel \cite{beyer2018site}, a highly configurable build system. BUILD files describe how binaries, libraries, and tests are built from source files. Most code is covered by our primary continuous integration system, ``TAP'' (Test Automation Platform \cite{beyer2018site}), and it is standard for TAP to gate releases. Bazel's builds and tests (including TAP) run on a shared set of machines called Forge. Forge's scale and cache enables Google to compile everything needed for a binary from scratch on every build, including fundamental dependencies like the Python interpreter.

\paragraph{Creating releases with Blueprints}

Google distinguishes between binaries and releases (named ``MPMs'' for the ``Midas Package Manager'' that stores them globally \cite{beyer2018site}). Releases are bundles of binaries and data that are ready to be deployed in \google's clusters, akin to a package in a Linux distribution. Releases are defined by Blueprint files, which are handwritten or managed by other systems that standardize releases and configurations. A system called Rapid \cite{beyer2016site} consumes Blueprints, runs CI tests, including TAP, and builds releases for server-side packages.

\paragraph{Running applications on Borg}

Borg is a custom cluster management service for the Google fleet that runs nearly all Google services \cite{43438}. Applications are deployed to Borg through configuration files that define the MPMs needed to run a service, runtime parameters, and scheduling constraints. MPMs can be rolled forward and backward safely because they are almost entirely hermetic.

\paragraph{Multiarch Support}

Borg was heterogeneous even before the Arm migration. Borg has had dozens of different types of CPUs over the years, and services run on machines that can be as much as ten years old. Unless an owner adds specific constraints, a job can be scheduled on any machine with an architecture-compatible binary. During development, engineers can request builds for multiple architectures (e.g., Arm, K8, Haswell) and expect multiarch MPMs. They can also request tests to run on each target hardware. At release time, owners of packages can also configure Blueprints to target one or more ISAs. Owners expect to get a mix of different kinds of machines with different performance profiles and, in some cases, ISAs.

\paragraph{Shifting down and Large Scale Changes (LSCs)}
\label{sec:overview:lscs}

Google has moved to a ``shift down'' approach to development \cite{GoogleCloudEvents2025ShiftDown}, where developers only focus on one level of the stack and other issues are abstracted and/or automated for them. For example, Bazel, TAP, and Rapid mean that developers do not generally need to worry about the specifics of CI. In addition, a healthy automated testing culture means everyone can change everyone else's code without frequent breakages. This enables Large Scale Changes (LSCs) \cite{winters2020software} which change code owned by many different teams and can affect thousands of files at once. In cases where an LSC is considered low risk, it can be approved centrally and submitted efficiently without asking individual teams. To get approval from many owners at once, Google has developed Rosie \cite{winters2020software} which allows engineers to create a very large commit and shard it into tens, hundreds, or thousands of smaller commits split up by owner.

\subsection{Life cycle of an ISA migration}

Moving an individual package from single arch (x86) to multiarch support requires several steps:

\begin{enumerate}
    \item \textbf{Test}: Fix tests (and builds) that break when run with the new ISA. Since anyone can build and test any code in our monorepo, it is easy to identify tests that break and require fixes.
    \item \textbf{Set up multiarch CI}: This requires modifying the corresponding Blueprint files to ensure that no additional regressions are introduced (often simultaneous with the next step).
    \item \textbf{Configure releases}: This modifies Blueprint files to make releases multiarch by default.
    \item \textbf{Roll out new binaries}: Run the multiarch packages on machines of the new ISA and assess performance and stability, addressing issues as needed.
    \item \textbf{Full production}: Allow production jobs to be scheduled on machines of the new ISA.
\end{enumerate}

\noindent While these steps are the same for all packages, the issues encountered within each step vary widely across applications and throughout the different phases of our ISA migration from x86 to Arm. Often, this involves performance-optimizing code for the new platform, which can happen in parallel to these steps. We observe parallels to Uber's reported porting workflow \cite{uber}.

\subsection{Phase 1: Large users}
\label{sec:overview:phase1}

We started our Arm migration with a small set of large users, such as Spanner \cite{39966}, BigQuery \cite{53334} and Bigtable \cite{10.5555/1267308.1267323}. These migrations were hands-on with a small team, and required weekly meetings and tracking bugs. Once tests passed, rollout was manual, with very careful performance and load testing, and gradually removing scheduling constraints on a per-job basis.

During this phase, a number of issues in these workloads were surfaced and addressed. Examples include:
\begin{enumerate*}[label=\emph{\arabic*}), itemjoin={{; }}]
\item Replacing x86-specific intrinsics
\item Replacing \texttt{long double}, which differs between x86 and Arm, with \texttt{absl::float128}
\item Brittle tests (e.g., due to exact floating point equality checks)
\item x86-specific flags
\item Memory ordering issues hidden by x86
\item Out-of-memory errors, often due to heap limits being tuned for x86
\item Multiarch MPMs exceeding the capacity limits of our infrastructure
\item Unsupported dependencies, and loading of unsupported dynamic libraries
\item Jobs not getting scheduled due to unsatisfiable scheduling constraints in Borg configurations.
\end{enumerate*}

This list was surprising to the teams involved---initially, there had been a perception that porting these large and mature codebases to Arm would be a herculean task, and that the very different toolchains would result in myriad difficulties. However, most issues involved simple changes or fixes, many of them in configuration files. At the same time, these changes were surprisingly pervasive, as evident by the large number of commits. It is therefore not the case that most software compiles and runs on Arm without modifications; it is that these modifications are of a different kind than expected initially. For example, it is not unusual that, for a given software package, almost \emph{nothing} builds initially, suggesting that large and pervasive changes are required. However, simple fixes to a number of shared dependencies often unblocks many of them at once.

\subsection{Phase 2: Everybody else}
\label{sec:overview:phase2}

To take full advantage of Arm in the data center, migrating only the largest workloads is insufficient. To make maximum use of available capacity, Borg needs to be able to schedule workloads flexibly across platforms, packing large and small users onto machines efficiently. If only a small subset of services can run on Arm, it will result in underutilization of those machines. We note that the distribution of workloads at Google is very flat: Although our top 50 users are very large, they only represent $\approx$ 60\% of running compute \cite{44271}. Addressing this long tail requires porting over 100,000 packages and billions of lines of code. This makes the Phase 1 approach of working directly with customer teams infeasible. In fact, even just talking to each team would be prohibitively expensive. The second phase of the x86 to Arm migration therefore focused on automating and scaling the migration of these workloads, while minimizing involvement from the teams themselves. It is this phase of \google's x86 to Arm migration that we mostly focus on. So far, we have ported about 30,000 packages, accounting for a significant portion of CPU cycles. We found that effectively making use of Arm hardware did not require porting all workloads.

\begin{figure}
    \centering
    \includegraphics[width=\textwidth]{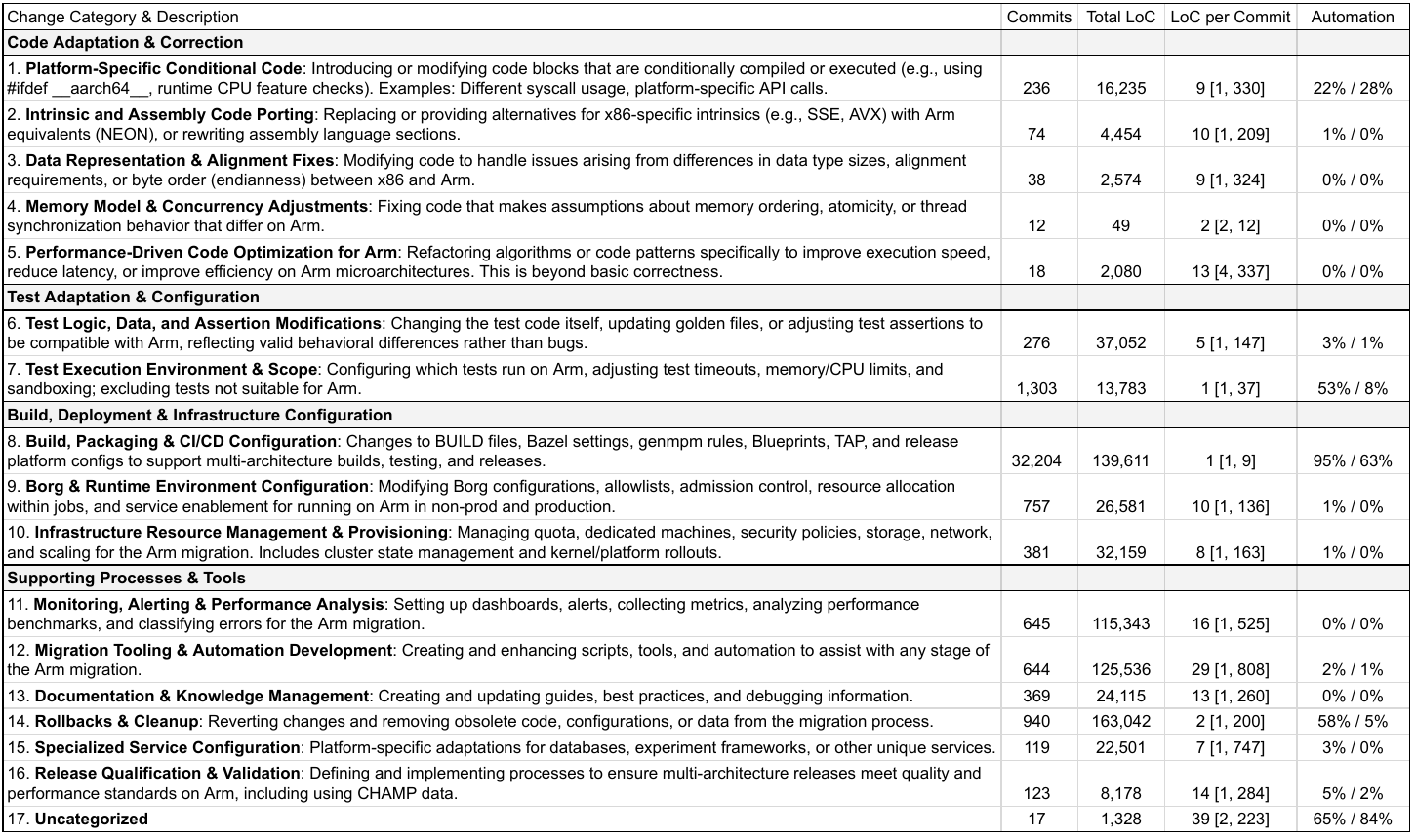}
    \caption{Categories of commits in \google's x86 to Arm migration. LoC per commit shows median and 90\% CI. Automation shows the fraction of commits/LoC generated using large-scale changes (Section~\ref{sec:automatability:tools}).}
    \label{fig:category_table}
\end{figure}

\section{Analyzing an ISA Migration}
\label{sec:analysis}

\begin{figure}
    \centering
    \includegraphics[width=\textwidth]{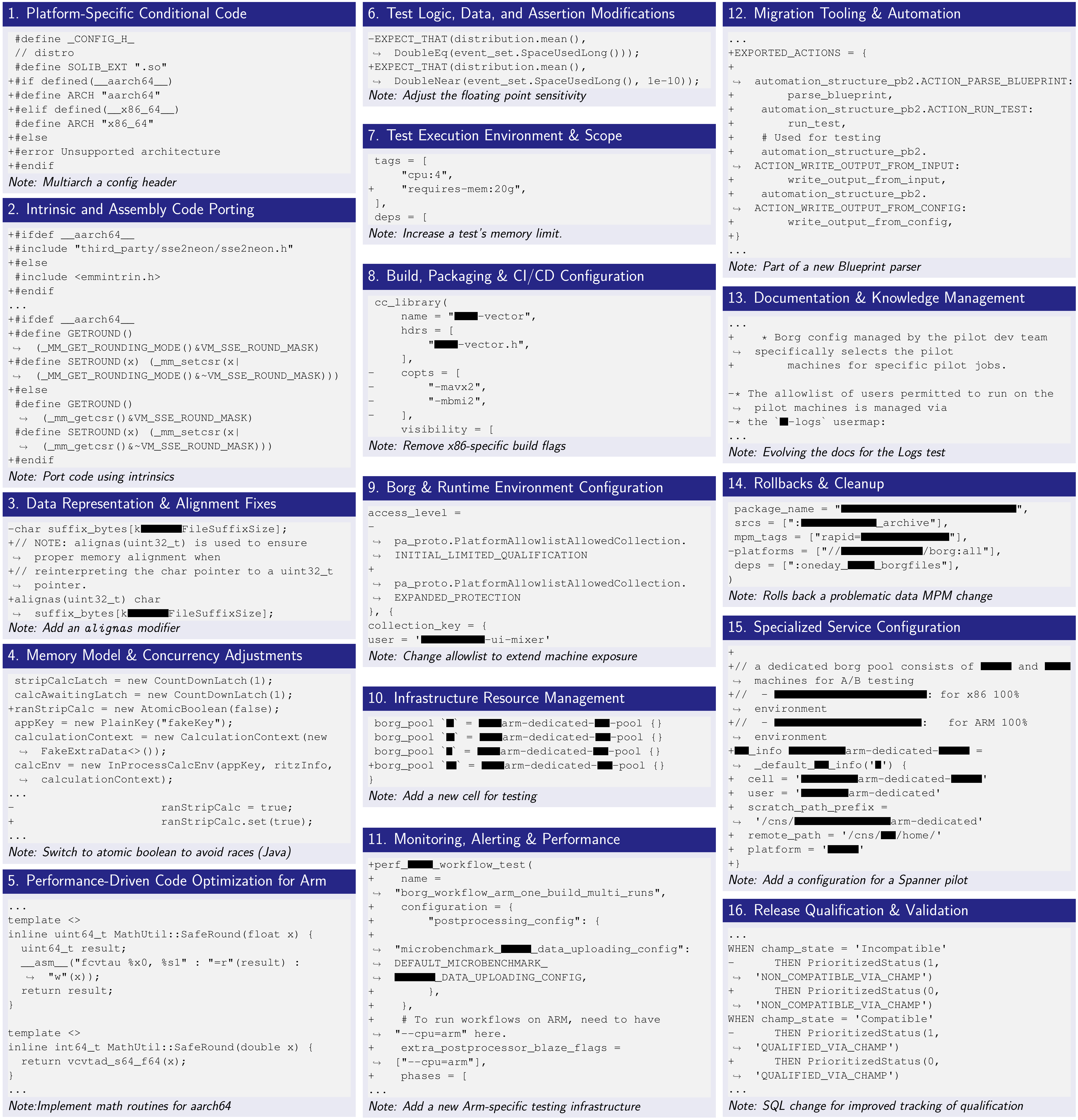}
    \caption{Specific code examples for each category.}
    \label{fig:code_examples}
\end{figure}

To fully understand what is involved in an ISA migration, we now analyze the full range of tasks involved in \google's x86 to Arm migration (RQ1). As a monorepo, any change -- be it to code, configurations or documentation -- is tracked as a commit in our repository's history. Further, these relevant commits were marked with a keyword that indicates they were part of this migration, allowing us to extract them after the fact. We thus identified a relevant set of \numberOfChanges commits.

Analyzing these commits manually would have been cost-prohibitive. We instead used a variant\footnote{We use Gemini 2.5 fine-tuned on a corpus of internal data, including code and documentation. This corpus may include material related to our Arm migration, but this represents a sufficiently small subset that recitation is not an issue.} of Gemini 2.5 Flash to analyze these commits at scale. We passed the commit messages and code diffs into the LLM’s 1M token context window in groups of 100 at a time. We prompted the model to pick a set of 20 categories for each batch. Then, we took all 400 $\times$ 20 categories and asked Gemini to consolidate them into 50. Further manual iteration over model outputs led to a final list of 16 categories (Figure~\ref{fig:category_table})\footnote{The descriptions in Figure~\ref{fig:category_table} are almost entirely the model's output, with some minimal edits to remove internal information.}. Once this list was finalized, we ran the model on all commits again and had it assign one of these 16 categories to each of them (as well as an additional ``Uncategorized'' category, which improved stability by catching outliers). Figure~\ref{fig:code_examples} shows examples of each category.

Commits fall into four overarching groups: 1) Code changes, 2) Test changes, 3) BUILD files and configurations, and 4) Supporting processes and tools. In total, our commits updated around 700K lines of code. While the vast majority (\categoryCommitFractionConfig) of commits are related to updating build or configuration files (Category 8), these commits account for only \categoryLoCFractionConfig of lines of codes updated. We also see a substantial number of lines (\categoryLoCFractionTooling) spent on migration tooling. A large portion of this work is only required once and can be reused in future ISA migrations. 

All categories contain a meaningful number of commits, supporting our claim that ISA migration is a multifaceted engineering challenge where no single type of task dominates. We also see that code related commits (Categories 1-5) only account for 1\% of commits and less than 4\% of lines of code, refuting the conventional wisdom \cite{825694} that code translation accounts for most of an ISA migration. In Section~\ref{sec:automatability}, we analyze how automatable these commits are.

\begin{figure}
    \centering
    \includegraphics[width=\textwidth]{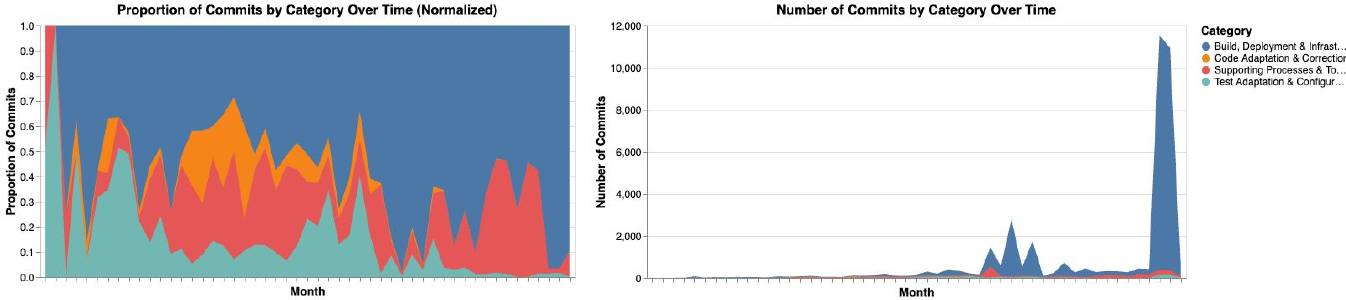}
    \caption{Categories of commits over time.}
    \label{fig:commits_over_time}
\end{figure}

We also analyze the timeline of our ISA migration (Figure~\ref{fig:commits_over_time}). We observe that at the start of the migration, most commits were in tooling and test adaptation, aligned with Phase 1 (Section~\ref{sec:overview:phase1}). Over time, a larger fraction of commits became around code adaptation, which can be seen as a phase when there is still a need to update code in common dependencies and address common issues in code and tests. Eventually, the fraction of these kinds of commits declines and in the final phase of the process (Section~\ref{sec:overview:phase2}), almost all commits are configuration files and supporting processes. We also observe that in this later phase, the number of merged commits rapidly increases, capturing the scale-up of the migration.

Finally, we want to understand how these different categories of commits differ from one another. We observe that median commits in most categories are less than 20 LoC, with many single-line commits. However, we also observe  very large individual commits that change 10,000+ LoC and skew the averages. We manually inspected these commits to understand their origin and found that these commits do not typically represent more work since they are conceptually similar to large numbers of simple, one-line commits. Overall, there are 19 commits that cumulatively account for 238,289 LoC (32.4\%) of total lines changed and that are trivial. Examples include:

\begin{itemize}
    \item Remove a porting tool once it was no longer used – 57K LoC (Category 12)
    \item Update a list of microbenchmark targets – 23K LoC (Category 11)
    \item Add several very large test vectors for a coverage tool – 15K LoC (Category 6)
\end{itemize}

\noindent In summary, we find that most commits related to migration are small, and that the largest commits often change very large lists or configurations and are not inherently complex. We also find that size alone does not measure difficulty. Finally, we observe that some of the commits (particularly in ``Supporting Processes \& Tools'') could likely be reused in a subsequent multiarch migration.

\section{Automating ISA Migrations}
\label{sec:automatability}

Now that we have established the tasks that are part of an ISA migration, we can explore how automatable each of these tasks is (RQ2) and how novel automation approaches can facilitate them.

\subsection{ISA Migration Automation at \google}
\label{sec:automatability:tools}

We already employ a number of automation tools at Google today that automate a large portion of the ISA migration process (\automatedCommitFraction of commits and \automatedCommitLoCFraction of LoCs).

\paragraph{Large-Scale Changes (LSCs)}

The key piece to ISA migration automation is Rosie (Section~\ref{sec:overview:lscs}), which allows us to programatically generate large numbers of commits and shepherd them through code review. This includes running affected TAP projects, requesting code reviews by code owners, and submitting each commit once all tests pass. We find that \totalAutomatedCommits of our commits were generated by Rosie, signaling automation. However, we note that these commits only account for \automatedCommitLoCFraction of lines of code, indicating that most of these commits are very small. Figure~\ref{fig:category_table} shows the fraction of each category that was generated using automated tools. For example, one major LSC adds the following line to Blueprint files of projects, configuring all of its tests and releases for Arm:

\begin{lstlisting}[language=code]
arm_variant_mode = ::blueprint::VariantMode::VARIANT_MODE_RELEASE,
\end{lstlisting}

\paragraph{Sanitizers \& Fuzzers} While not limited to ISA migrations, fuzzers and LLVM sanitizers such as AddressSanitizer \cite{10.5555/2342821.2342849}, MemorySanitizer \cite{43308} and ThreadSanitizer \cite{35604} are key enablers of our migration. Even before Arm adoption, Google routinely ran all TAP tests with these tools enabled, which turn latent errors such as a memory corruption, memory leak, or race condition into a debuggable fault. Application owners regularly triage and fix these faults, causing us to sidestep many common differences in execution between x86 and Arm (e.g., a data race may be hidden by x86's TSO memory model). Catching these kinds of issues ahead of time avoids debugging non-deterministic and hard-to-debug behavior when recompiling to a new ISA.

\paragraph{Continuous Health Monitoring Platform (CHAMP)} The final step in our automation is CHAMP, which assesses Arm-built applications on Arm server hardware. It continuously monitors health metrics to detect whether behavior differs from x86 instances of the job (e.g., significantly higher RPC error rates or crashes). If so, it automatically marks the job as ineligible for Arm, files a bug for its owners to follow up, and automatically retries in 30 days. It scales up the fraction of Arm instances of the application task by task, job by job, and cell by cell following Google's production principles to limit SLO risk. CHAMP is not needed for new microarchitecture deployments (either x86 or Arm), as the behavior, performance differences, and associated issues are relatively minor. However, auto-qualification of Arm binaries was necessary due to the increased incidence of issues.

Using CHAMP, it is no longer necessary to manually shepherd every binary through qualification. Instead, after updating project configurations to build Arm releases, this process is now automatic.

\subsection{Reliability of the Automation Approach}

Combined, these tools allow for a mostly-automated approach where LSCs enable Arm for different builds and releases, which are then automatically qualified using CHAMP. To understand the stability of this approach, we analyze LSCs targeting a standardized release management system. These LSCs modified release configurations to bring this system's percentage of Arm-qualified applications from 4.8\% to 59.6\%. The rate of applications that were rolled back in early testing was 1.8\% (which dropped to 1\% after fixing bugs), and less than 0.8\% in the final phase.

Early in the migration, after $\approx$ 300 MPMs, we had a 5\% refusal rate (code owners deciding not to migrate). During scale up, this dropped to 0.6\% after $\approx$ 600 additional MPMs. In the final phase, the commits were globally approved, with no refusal. We found that acceptance rate was strongly influenced by careful workload targeting, users gaining trust in the automation, and messaging that anticipated worries and objections.

\section{Automation of ISA Migrations with AI}

While LSCs and CHAMP automate a large part of the porting process, there are limits to this approach. They can edit build and configuration files, as well as automatically qualify Arm binaries for deployment. However, standard LSCs are fixed-function pipelines. They are not flexible to respond to unexpected errors or other issues that occur at any stage of the process, be it during testing or in production.

Modern generative AI techniques represent an opportunity to automate the remainder of the ISA migration process. We built an agent called \emph{CogniPort} which aims to close this gap. CogniPort operates on build and test errors. If an Arm binary does not build or a test fails at any point in the process, the agent steps in and aims to fix the problem automatically.

\begin{figure}
    \includegraphics[width=0.6\textwidth]{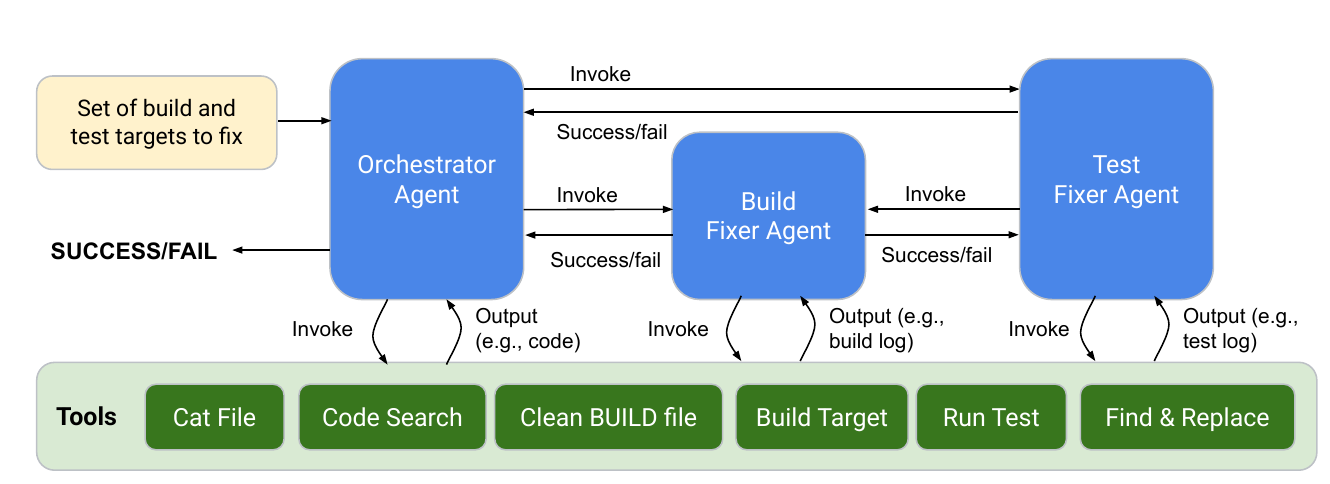}
    \includegraphics[width=0.35\textwidth]{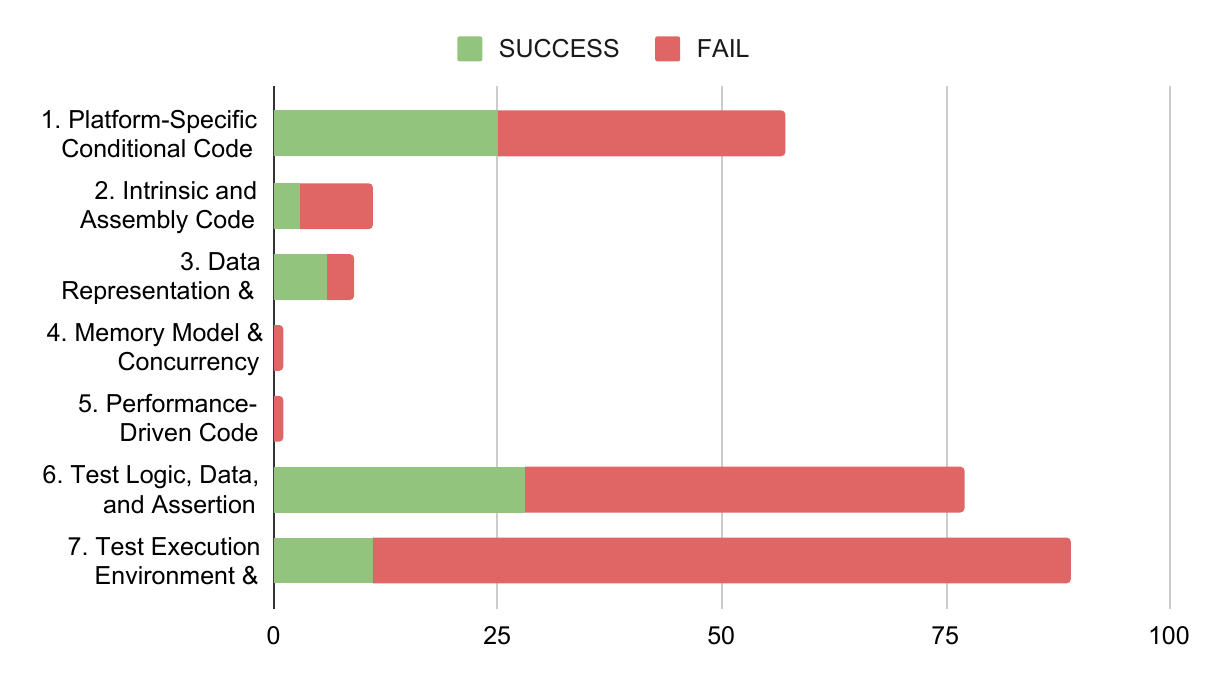}
    \caption{Agentic flow and the success rate of the agent.}
    \label{fig:agent}
\end{figure}

The agent consists of three nested agentic loops (Figure \ref{fig:agent}). Each loop executes the LLM from Section~\ref{sec:analysis} to perform one step of reasoning, followed by a tool invocation---i.e., a function call (the loop terminates once the agent emits a special `finish' call). This tool executes, and its outputs are attached to the agent's context. For example, there are tools for building and returning the build log, running test(s) and returning the test log, and running a tool that fixes errors in BUILD files. The agent also has tools to search through code and make modifications.

The outermost agent loop is an \emph{orchestrator} that repeatedly calls the \emph{build fixer agent} and/or the \emph{test fixer agent} depending on the state of the workspace. The build fixer agent tries to build a given target and makes modifications to files until the target builds successfully. The test fixer agent tries to run a given test and makes modifications until the test passes. In both cases, the agent can time out via a step limit or give up early by calling `finish`.

To evaluate the agent, we take historic commits from our data set, revert them and then evaluate whether the agent is able to fix them. We note that not all of our categories are suitable for this approach---it only applies to Code \& Test Adaptation (categories 1-8). To evaluate the agent, we further narrow down our data set by only picking commits that can be cleanly reverted and that have identifiable build or test targets. This results in a benchmark set of 245 commits.

We see that the overall success rate is 30\%, with test fixes, platform-specific conditionals, and data representation fixes having the highest success rates. Memory model, test execution environment, and performances fixes are most difficult (though based on a very small sample size). Overall, however, this indicates that AI achieves a reasonably high success rate. We note that we consider these results \emph{directional}: While we analyzed an arbitrarily chosen subset of outputs to ensure the validity of results and confirm that we are not observing recitation from training, the evaluation is not perfect and may miss cases where, e.g., a fix is incorrect but not caught by a test or where there is other information leakage (e.g., a subsequent commit made the original fix easier or reverting only the commit itself makes it easier to root cause an issue than if the entire fix was reverted).
\section{Discussion \& Research Challenges}

\begin{figure}
    \includegraphics[width=\textwidth]{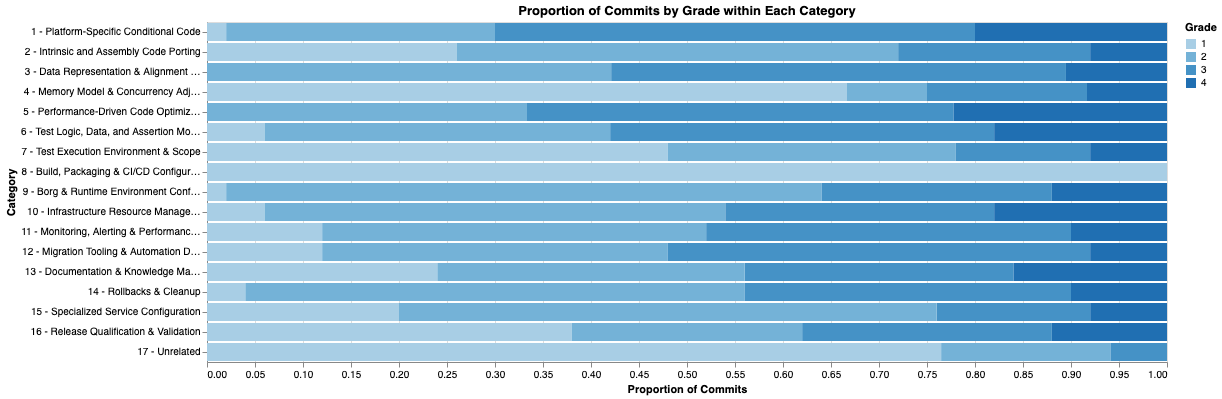}
    \caption{AI-assessed automatability of each category (1 = trivial, 5 = probably unsolvable), as well as actual fraction of commits and LoCs in each category that were generated using automated tools.}
    \label{fig:automatability-scores}
\end{figure}

Overall, we see that ISA migrations involve a wide range of different tasks. Many of these tasks are highly automatable. BUILD file and configuration changes are almost fully automatable, while code changes and tests are partially automatable with AI. In addition, there are many changes in areas like build and test infrastructure, spinning up a new hardware platform whether or not it is a new ISA (e.g., categories 11 and 15), or deprecating old code (category 14) that needed generalizations for multiarch support, but these scale across many architectures and will not be required again.

This refutes the conventional wisdom that the main challenge of ISA migration is translation, and also that migrating a large software ecosystem to a new ISA is a prohibitively large amount of work, particularly with modern AI. This raises the question: What are the remaining challenges, and are there research opportunities in closing the remaining gaps (RQ3)?

To answer this, we analyze our data set of commits to identify changes that are challenging to automate. Once again, we use AI. We used our LLM to assess how automatable different categories are by sampling up to 50 commits in each category and using an LLM to grade them (1 = trivially automatable, 5 = difficult, even for advanced AI). This is not a conclusive result, but directionally tells us how difficult each category is, according to the LLM itself (Figure~\ref{fig:automatability-scores}). By manually inspecting the outputs, a picture emerges of which changes remain the most challenging.

First, we find that the LLM confirms the categories that are already automatable today---i.e., BUILD and configuration files, test execution environment---as automatable. Second, categories 1-7 (code \& test adaptation) stand out as problems that have a significant fraction of commits ranked as 3 and 4---indicating problems that are hard but not impossible. Examples of these include:

\begin{itemize}
  \item \textbf{ISA-Specific Vector Code}: Writing ISA-specific, performant vector code is a hard problem, and is actively investigated by the research community \cite{10.1145/3696443.3708929}. While current AI can translate simple routines, generating complex kernels exposes a complex search space that goes beyond a simple build-repair loop.
  \item \textbf{Deep Performance Optimizations}: Performance optimizations sometime require major refactors, algorithmic changes and intrinsics. There is significant existing work towards applying LLMs to these problems. At Google, we have a system called ECO that is used for automating some of these optimizations \cite{lin2025ecollmdrivenefficientcode}, including for code running on Arm.
  \item \textbf{Difficult Corner Cases}: We saw a number of corner cases that require obscure knowledge beyond the code itself. For example, we saw commits that worked around an Arm compiler bug, addressed a hash function behaving differently on Arm and x86, and fixed bugs that were unrelated to the Arm migration but were not previously triggered. It is plausible that an agent that can search documentation and the wider web could perform better on these.
  \item \textbf{Performance Tuning}: Hyperparameters and feedback directed optimization (FDO) profiles sometimes have to be regenerated for a new platform. This is potentially automatable, but requires an agent to be able to run workloads and perform performance measurement.
\end{itemize}

\noindent Meanwhile, we also found a number of examples where it is difficult to tell whether they are automatable using AI or whether they fundamentally require human involvement:

\begin{itemize}
  \item \textbf{Multiarch tooling}: A significant portion of this work includes implementing the automation itself (e.g., CHAMP), as well as simulation tools and dashboards. While AI can help in the development flow of these components, many commits involve addressing feature requests by users, and thus require human involvement. On the other hand, this work only needs to be done once and is not required for future ISA migrations.
  \item \textbf{Resource provisioning}: These are changes to configuration that follow human engineers installing hardware in data centers and subsequent management by Borg. They should not be performed by AI, and are being obviated via improvements to Borg and adjacent systems.
  \item \textbf{Documentation}: LLMs have already proven useful at generating some kinds of documentation \cite{10.1145/3664646.3664765}, and future LLMs may improve the ability of maintaining it and generating user-focused narrative documentation (and chat agents) that require context beyond code.
\end{itemize}

\noindent Taken together, this demonstrates that there are opportunities for further closing the gap and that future ISA migrations may require a limited amount of manual work, mostly focused on making new hardware available and adding the new ISA to the automated multiarch tooling.
\section{Conclusion}

By analyzing a large-scale ISA migration at \google, we refute several long-standing assumptions about ISA migrations. First, code translation is only a small portion of the ISA migration, and mainly focuses on intrinsics and vector code. Second, merely recompiling available code is not sufficient. Third, the tasks required for an ISA migration are multifaceted, with no single task dominating. Fourth, many of these tasks are highly automatable, particularly using modern AI techniques. Finally, even with AI, there remain a number of challenges that currently require human involvement and represent opportunities for future work on AI for ISA migration.

\section*{Acknowledgments}

We thank Arvind Sundararajan, Dushyant Acharya, Ahmed Alansary, Shelah Ameli, Owen Anderson, Sterling Augustine, Sushmita Azad, Nupur Baghel, Antoine Baudoux, Patrick Bellasi, Vincent Belliard, Kyle Berman, Paul Bethe, Gopu Bhaskar, Raymond 'Princess Sparklefists' Blum, Marshall Bockrath, Harsha Vardhan Bonthalala, Lexi Bromfield, Jean-Luc Brouillet, Eric Burnett, Marcelo Cataldo, John Cater, Kristine Chen, David Cheng, Ilya Cherny, Saahithi Chillara, Rob Chilton, Chandrakanth Chittappa, Brian Chiu, Daniele Codecasa, Eduardo Colaço, Pavithra Dankanikote, Nicolo Davis, Rumeet Dhindsa, Zhuoran Diao, Bartosz Dolecki, Ian Dolzhanskii, Pat Doyle, Elian Dumitru, Ali Esmaeeli, Samuel Foss, Ákos Frohner, Neeharika Gonuguntla, Shruti Gorappa, Russell Gottfried, Manoj Gupta, Benjamin Gwin, Yanyan Han, Jitendra Harlalka, Milad Hashemi, Tim Henderson, Daisy Hollman, Jung Woo Hong, Jiawei Huang, Jin Huang, Talha Imran, Victoria Juan, Pranav Kant, Lera Kharatyan, Joonsung Kim, Danial Klimkin, Sree Kodakara, Avi Kondareddy, Danila Kutenin, Gregory Kwok, Pavel Labath, Leon Lee, Sungkwang Lee, Li Li, Sha Li, Xu Li, Zhongqi Li, Jianyi Liang, Kevin Liston, Haiming Liu, Li Liu, David Lo, Sean Luchen, Albert Ma, Laura Macaddino, Anthony Mai, Jennifer Mansur, Simon Marchuk, David Margolin, Alexander Midlash, Dominic Mitchell, Karthik Mohan, Albert Morgese, Maksym Motornyy, Katherine Nadell, Ngan Nguyen, Denis Nikitin, Stoyan Nikolov, Nicolas Noble, Chester Obi, Andri Orap, Habib Pagarkar, Dasha Patroucheva, Sabuj Pattanayek, Vaishakhi Pilankar, Saranyan Vangal Rajagopalan, Daniel Rall, Majid Rasouli, Salonik Resch, Alberto Rojas, Annie Rong, Jesse Rosenstock, Michal Sapinski, Yashnarendra Saraf, Aaron Schooley, Alexander Schrepfer, Manish Shah, Alexander Shaposhnikov, Stan Shebs, Guangyu Shi, Oscar Shi, Santanu Sinha, Anna Sjövall, Ben Smith, Justin Smith, Jairaj Solanke, Fangrui Song, Raman Subramanian, Xenia Tay, Toni Thompson, Dima Tsumarau, Cassie(Yitong) Wang, Tommy Wang, Shu-Chun Weng, DJ Whang, Hailong Xiao, Andy Xu, and Jason Yuan for their contributions to Google's Arm porting efforts and the work described herein.


\end{document}